\documentstyle[epsf]{mn}
\input{psfig.tex}

\def\VEV#1{\left\langle #1\right\rangle}
\def\vecw{{\vec w}}
\def\veckappa{{\vec\kappa}}
\long\def\comment#1{}
\def\VEV#1{\left\langle #1\right\rangle}
\def\fun#1#2{\lower3.6pt\vbox{\baselineskip0pt\lineskip.9pt
  \ialign{$\mathsurround=0pt#1\hfil##\hfil$\crcr#2\crcr\sim\crcr}}}
\def\lap{\mathrel{\mathpalette\fun <}}
\def\gap{\mathrel{\mathpalette\fun >}}

\begin{document}

\title{Theory and Statistics of Weak Lensing from Large-Scale
Mass Inhomogeneities}

\author[Kamionkowski et al.]{Marc Kamionkowski$^{1,2}$, Arif
Babul$^{3,4}$, Catherine M. Cress$^{2,5}$, Alexandre
Refregier$^{1,2,6}$ \\
$^1$Department of Physics, Columbia University, 538 West 120th
     Street, New York, NY 10027 U.S.A. \\
$^2$Columbia Astrophysics Laboratory, 538 West 120th St., New
     York, NY 10027 U.S.A. \\
$^3$Department of Physics \& Astronomy, University of
     Victoria, P. O. Box 3055, Victoria, BC  V8W 3P6 Canada \\
$^4$Department of Physics, New York University, 4 Washington
     Place, New York 10003 U.S.A. \\
$^5$Department of Astronomy, Columbia University, 538 West 120th
     Street, New York, NY 10027 U.S.A. \\
$^6$Department of Astrophysical Sciences, Peyton Hall, Princeton
     University, Princeton, NJ 08544 U.S.A.}

\maketitle

\begin{abstract}
Weak lensing by large-scale mass inhomogeneities in the Universe
induces correlations in the observed ellipticities of distant
sources.  We first review the harmonic analysis and
statistics required of these correlations and discuss
calculations for the predicted signal.  We consider the
ellipticity correlation function, the mean-square ellipticity,
the ellipticity power spectrum, and a global maximum-likelihood
analysis to isolate a weak-lensing signal from the data.  
Estimates for the sensitivity of a survey of a given area, 
surface density, and mean intrinsic source ellipticity are
presented.  We then apply our results to the {\sl FIRST}
radio-source survey.   We predict an rms ellipticity of roughly
0.011 in $1^\circ \times 1^\circ$ pixels and 0.018 in
$20' \times 20'$ pixels if the power spectrum is
normalized to $\sigma_8 \Omega^{0.53} = 0.6$, as indicated by
the cluster abundance.  The signal is significantly larger
in some models if the power spectrum is normalized instead to
the {\sl COBE} anisotropy.  The uncertainty in the predictions
{}from imprecise knowledge of the {\sl FIRST} redshift
distribution is about 25\% in the rms ellipticity.  We show
that {\sl FIRST} should be able to make a statistically
significant detection of a weak-lensing signal for
cluster-abundance--normalized power spectra.
\end{abstract}

\begin{keywords}
cosmology: theory - gravitational lensing -
large-scale structure
\end{keywords}

\maketitle


\section{INTRODUCTION}

It has been proposed that the effects of weak
lensing on distant sources could shed light on the large-scale
distribution of mass in the Universe
\cite{gun67,mir91,bla91,kai92,bar92,kai96,vil96,ste97,jai97}.
Mass inhomogeneities
along the line of sight to distant sources will induce
distortions in the images of these sources; thus, correlations
of the ellipticities of distant sources provides a probe of the
correlation of mass along the line of sight.  In this way, the
power spectrum for the {\it mass} (rather than light)
distribution in the Universe can be probed.  

This technique requires (i) a sample of sources which are
distant, so there is a large line of sight over which the
lensing signal can accrue; (ii) good angular resolution, so that
the ellipticities of sources can be determined; and (iii) a
large enough sample of sources so that the noise provided by the
intrinsic ellipticities of sources can be overcome.  For
example, the VLA {\sl FIRST} radio survey meets all of these
criteria \cite{bec95,whi97}.  Upon
completion, the survey will cover 10,000 square degrees of the
North Galactic cap.  There are $\sim40$ sources per
square degree with resolved structure on scales from $2-30''$
at the survey resolution of $5''$, and the mean redshift of these
sources is of order unity.  Several systematic effects can
produce spurious ellipticity correlations in {\sl FIRST}, and
therefore be mistaken for a weak-lensing signal. However, the
most serious systematic effects are understood and can be
corrected for \cite{ref97,refetal97}.  

When looking for the effects of weak lensing on galaxies behind
a cluster, one requires tens of thousands of galaxies per square
arcminute to overcome
the Poisson noise from intrinsic galactic ellipticities and
thereby map the shear field.  With this in mind, it may at first
seem hopeless to detect the effects of weak lensing in a
sparsely-sampled survey such as {\sl FIRST} with $<100$ resolved
sources per square degree.  However, for weak lensing from
large-scale structure, we are interested in the correlation of
ellipticities of pairs of sources with some fixed angular
separation; we do not necessarily need to map the shear field.
For this, the relevant quantity is not the density
of sources, but the {\it total} number of pairs of sources in
the survey with some fixed angular separation.  In other words,
the sensitivity to the mean-square ellipticity averaged over
regions of some fixed size on the sky is improved with a large
area [cf., the discussion of sparse sampling in Kaiser (1996)].

In this paper, we review the theory of ellipticity correlations
{}from weak gravitational lensing.  We discuss
statistical techniques which can be used to isolate a signal in
the data.  We also estimate the amplitude which may be
detectable with a survey which covers a given fraction of the
sky with a given number of resolved sources per square degree
and a given mean intrinsic source ellipticity.  
We then calculate the
predicted weak-lensing correlations in the {\sl FIRST} radio
survey for the canonical cold-dark-matter model as well as for
several viable variants.  We discuss the uncertainties in the
predictions which arise from imprecise knowledge of the redshift
distribution of {\sl FIRST} sources, and consider the
detectability of the signal.  Refregier \&
Brown (1997) show how spatial noise correlation, one of the most
serious systematic effects for {\sl FIRST}, affects weak-lensing
measurements and can be corrected for.  A subsequent paper will
present the results of the search for weak lensing in {\sl
FIRST} \cite{refetal97}.

\section{ELLIPTICITY CORRELATIONS FROM WEAK LENSING}

\subsection{Description of the Shear Field}

Weak lensing will induce a stretching of images on the sky at
position $\vec\theta=(\theta_x,\theta_y)$ described by the shear
field, a symmetric, trace-free $2\times2$ tensor field:
\begin{equation}
     \gamma_{\alpha\beta}(\vec\theta) = \left( \begin{array}{cc}
     \epsilon_+(\vec\theta) & \epsilon_\times(\vec\theta)  \\
     \epsilon_\times(\vec\theta) & -\epsilon_+(\vec\theta)  \\
     \end{array} \right).
\end{equation}
Here, $\epsilon_+$ is the stretching in the
$\hat\theta_x - \hat\theta_y$ directions, and $\epsilon_\times$
is the stretching along axes rotated by $45^\circ$.
Alternatively, the shear field can be written as a shear ``vector,''
\begin{equation}
     \gamma_\alpha = (\epsilon_+,\epsilon_\times)= \gamma(\cos
     2\alpha,\sin2\alpha),
\end{equation}
although this ordered pair does not transform as a vector.
The deformation is also sometimes written as a complex
ellipticity $p$; the two components of the polarization
``vector'' are the real and imaginary parts of the complex
ellipticity, 
\begin{eqnarray}
     p&=&{(a^2 - b^2)\over(a^2+b^2)} e^{2i\alpha} = |p|
     e^{2i\alpha} \nonumber \\
     &=& (\epsilon_+^2 + \epsilon_\times^2)^{1/2} e^{2i\alpha} =
     \epsilon_+ + i \epsilon_\times,
\end{eqnarray}
where $a$ and $b$ are the principle moments.  We see that
$\gamma = |p| = (\epsilon_+^2 +\epsilon_\times^2)^{1/2} = (a^2 -
b^2) /(a^2+b^2)$ and $\tan 2\alpha= \epsilon_\times/\epsilon_+$.

As pointed out by Stebbins (1997), the shear tensor field
$\gamma_{\alpha\beta}(\vec\theta)$, can be decomposed into a
``gradient'' or curl-free component (what Stebbins calls
the scalar part) and a ``curl'' (or pseudoscalar) component,
just as a two-dimensional vector field can be decomposed into
curl and curl-free parts. In other words, the shear tensor can
be written in terms of two scalar functions
$\gamma_g(\vec\theta)$ and $\gamma_c(\vec\theta)$,
\begin{equation}
     \nabla^2 \gamma_g = \partial_\alpha \partial_\beta \gamma_{\alpha\beta},
     \quad 
     \nabla^2 \gamma_c = \epsilon_{\alpha\gamma} \partial_\beta
     \partial_\gamma \gamma_{\alpha\beta},
\end{equation}
where $\epsilon_{\alpha\beta}$ is the antisymmetric tensor.
Specification of $\gamma_g(\vec\theta)$ and $\gamma_c(\vec\theta)$
is equivalent to specification of $\epsilon_+(\vec\theta)$ and
$\epsilon_\times(\vec\theta)$.  The quantities $\gamma_g$ and
$\gamma_c$ are the gradient and curl components of the
ellipticity field.  

Density perturbations (mass inhomogeneities) produce only scalar
perturbations to the spacetime metric, so they have no handedness
and can therefore produce no curl.  Gravitational waves have a
handedness and may induce a nonzero $\gamma_c$.  However, the
weak-lensing signal from gravitational waves is expected to be
extremely small \cite{ste97}. Vector modes could also produce
a curl, but, like tensor modes, they are 
negligible. Therefore, we expect that
$\gamma_c=0$, and only $\gamma_g$ should be nonzero.  This 
allows one to isolate the effect of weak lensing and to
check for non-lensing ellipticity correlations.

Throughout, we approximate the region of sky surveyed as a flat
surface.  At first this might seem inappropriate since the {\sl FIRST}
survey covers a good fraction of the sky.  However, weak-lensing
ellipticity correlations should be most significant only at smaller
angular separations, so the inaccuracies which arise from
approximating the sky as a flat surface should be small.
Furthermore, we are primarily interested here in establishing
the existence of an ellipticity correlation.  With future more
sensitive data, it will be necessary to account for the
curvature of the sky \cite{ste97}.  

Given the shear tensor $\gamma_{\alpha\beta}(\theta)$, the functions
$\gamma_g$ and $\gamma_c$ can be constructed only with a
Fourier transform.  Writing
\begin{eqnarray}
     \gamma_{\alpha\beta}(\vec\theta) &=& \int\, {d^2 \vec\kappa \over
     (2\pi)^2}\, \tilde 
     \gamma_{\alpha\beta}(\vec\kappa)\, e^{-i \vec\kappa \cdot
     \vec\theta}, \nonumber \\
     \tilde\gamma_{\alpha\beta}(\vec\kappa) &=& \int\, d^2
     \vec\theta \,
     \gamma_{\alpha\beta}(\vec\theta)\, e^{i \vec\kappa \cdot \vec\theta},
\end{eqnarray}
where the tilde denotes the Fourier transform (and similarly for
the other quantities), we get
\begin{equation}
     \tilde \gamma_g(\vec\kappa) = { (\kappa_x^2 -\kappa_y^2) \tilde
     \epsilon_+(\vec\kappa) + 2\kappa_x \kappa_y \tilde
     \epsilon_\times(\vec\kappa)  \over
     \kappa_x^2 + \kappa_y^2},
\end{equation}
\begin{equation}
     \tilde \gamma_c(\vec\kappa) = { 2\kappa_x \kappa_y \tilde
     \epsilon_+(\vec\kappa)  -
     (\kappa_x^2 -\kappa_y^2) \tilde \epsilon_\times(\vec\kappa) \over
     \kappa_x^2 + \kappa_y^2}.
\end{equation}
The functions $\gamma_g(\vec\theta)$ and $\gamma_c(\vec\theta)$
can then be recovered through the inverse Fourier transformation.
Since $\gamma_{\alpha\beta}$ is a real tensor, $\tilde
\gamma_{\alpha\beta}^*(\vec\kappa) 
= \tilde \gamma_{\alpha\beta}(-\vec\kappa)$, and similarly for $\tilde
\gamma_g$ and $\tilde \gamma_c$.

\subsection{Power Spectra}

Statistical homogeneity and isotropy
guarantee that the two
sets of Fourier coefficients, $\tilde\gamma_g$ and
$\tilde\gamma_c$, have expectation values,
\begin{eqnarray}
    \VEV{ \tilde\gamma_g^*(\vec\kappa)
    \tilde\gamma_g(\vec\kappa')} &=& (2\pi)^2
    \delta(\vec\kappa -\vec\kappa') P_{gg}(\kappa), \nonumber \\
    \VEV{ \tilde\gamma_c^*(\vec\kappa) \tilde\gamma_c(\vec\kappa')} &=& (2\pi)^2
    \delta(\vec\kappa -\vec\kappa') P_{cc}(\kappa), \nonumber \\
    \VEV{ \gamma_g^*(\vec\kappa) \gamma_c(\vec\kappa')} &=& (2\pi)^2
    \delta(\vec\kappa -\vec\kappa') P_{gc}(\kappa).
\end{eqnarray}
The power spectrum $P_{gg}(\kappa)$ is precisely Kaiser's (1992)
ellipticity power spectrum $P_\epsilon(\kappa)$.
The second power spectrum $P_{cc}(\kappa)$ will be effectively
zero because weak lensing from gravitational waves is extremely
small.  The third, $P_{gc}(\kappa)$, must be identically zero
since it is parity violating---that is, this power spectrum
changes sign under the change of coordinates ${\bf \hat x}
\rightarrow - {\bf \hat x}$.  Since these latter two power
spectra are zero, they can be used to look for non-lensing
artifacts in the data.

The mean-square gradient component of the ellipticity is
\begin{equation}
     \VEV{ \gamma_g^2} = \int {d^2 \vec\kappa \over (2\pi)^2}
     P_{gg}(\kappa).
\end{equation}
Since the mean-square curl component of the ellipticity is zero,
$\VEV{ \gamma_g^2}$ is also the mean-square total ellipticity.

Realistically, the mean-square ellipticity cannot be measured.
The actual measured quantity is the mean-square ellipticity
smoothed with some window function.  Suppose we estimate
the shear field at position $\theta$ by averaging over all
ellipticities, e.g., in a square $\theta_p \times \theta_p$
pixel centered at $\theta$.  In that case, we are probing a
smoothed shear field,
\begin{equation}
     \epsilon_+^s(\vec\theta) = \int\, d^2\vec\alpha\,
     W(\vec\alpha) \, \epsilon_+(\vec\theta+\vec\alpha),
\end{equation}
where $W(\vec\alpha)$ is the window function (e.g., for square
pixels, constant inside the pixel, zero outside, and normalized
to unity), and similarly for $\epsilon_\times^s$.  All prior
results for the unsmoothed field are generalized to the smoothed
field as long as we replace $\tilde\epsilon_+(\vec\kappa)
\rightarrow \tilde\epsilon_+^s(\vec\kappa) =
\tilde\epsilon_+(\vec\kappa) \widetilde W(\vec\kappa)$, where
$\widetilde W(\vec\kappa)$ is the Fourier transform of
$W(\vec\theta)$.  The mean-square smoothed ellipticity is then
\begin{equation}
     \VEV{ (\gamma_g^s)^2} = \int \, {d^2 \vec\kappa \over
     (2\pi)^2} \, P_{gg}(\kappa)\, |\widetilde W(\vec\kappa)|^2.
\label{eq:smoothedvariance}
\end{equation}
In the following, we also use the shorthand $P_{gg}^s(\kappa)
\equiv P_{gg}(\kappa)|\widetilde W(\veckappa)|^2$ for the smoothed
power spectrum.  Although the mean-square ellipticity gives a
simple indication of the magnitude of the weak-lensing signal,
one can obtain a much more sensitive probe of a signal by taking
advantage of the information provided by the complete power
spectra (or equivalently, correlation functions), as discussed
further below.

\subsection{Correlation Functions}

There are three independent two-point ellipticity correlation
functions that can be constructed in configuration space from
the three power spectra.
Since the components ($\epsilon_+$ and $\epsilon_\times$) of the
shear tensor are defined with respect to some set of axes on the
sky and transform under rotations, correlation
functions of these quantities will depend on the relative
orientation of the two points being correlated as well as the
separation. However, correlation functions which are independent
of the coordinate system can be constructed \cite{ste97} in
analogy with those needed for CMB polarization correlations
\cite{kam97}.   To do so, we define
correlation functions of ellipticities $\epsilon_+^r$ and
$\epsilon_\times^r$ measured with respect to 
axes which are parallel and perpendicular to the line connecting
the two points being correlated.  To be explicit, suppose the
first point is $\vec\theta_1=(\theta_{1x},\theta_{1y})$ and the
second is $\vec\theta_2=(\theta_{2x},\theta_{2y})$.  Then we must
rotate the axes by an angle $\phi = \arctan[(\theta_{2y} -
\theta_{1y})/(\theta_{2x}-\theta_{1x})]$ to align the rotated
$x$ axis with the line connecting the two points.  Under this
rotation, we get
\begin{eqnarray}
     \epsilon_+^r & =& \epsilon_+ \cos 2 \phi + \epsilon_\times
     \sin 2\phi, \\
     \epsilon_\times^r & =& -\epsilon_+ \sin 2 \phi + \epsilon_\times
     \cos 2\phi,
\end{eqnarray}
at both points.  The $2\phi$ enters since the ellipticity is
unchanged under a rotation by $90^\circ$.  We can then construct
three correlation
functions, $\VEV{\epsilon_+^r \epsilon_+^r}$,
$\VEV{\epsilon_\times^r \epsilon_\times^r}$,  $\VEV{\epsilon_+^r
\epsilon_\times^r}$, from the rotated components.  Although
$\epsilon_+^r$ is invariant under reflection along the line
connecting the two points being correlated, $\epsilon_\times^r$ 
changes sign.  Therefore, parity invariance demands that
$\VEV{\epsilon_+^r \epsilon_\times^r}=0$.  Statistically
significant deviations from zero can be due only to systematic
errors in the data.

By setting $\phi=0$ in Kaiser's Eq. (2.3.1), we identify
$\VEV{\epsilon_+^r(\vec\theta_0) \epsilon_+^r(\vec\theta_0+\vec\theta)} =
C_1(\theta)$ and 
$\VEV{\epsilon_\times^r(\vec\theta_0)
\epsilon_\times^r(\vec\theta_0+\vec\theta)} = C_2(\theta)$.  We 
also verify that $\VEV{\epsilon_+^r(\vec\theta_0)
\epsilon_\times^r(\vec\theta_0+\vec\theta)} = 0$.  In
analogy with correlation functions of Stokes parameters of the
cosmic microwave background \cite{kam97}, we can
write the correlation functions (for any $\phi$) in terms of the
power spectra as
\begin{eqnarray}
     C_1(\theta)+C_2(\theta) &=& \int_0^\infty {\kappa\, d\kappa \over 2\pi}
     [P_{gg}(\kappa) + P_{cc}(\kappa)] J_0(\kappa\theta)
      \\
     C_1(\theta)-C_2(\theta) &=& \int_0^\infty {\kappa\, d\kappa \over 2\pi}
     [P_{gg}(\kappa) - P_{cc}(\kappa)] J_4(\kappa\theta).
\end{eqnarray}
In terms of correlation functions of rotated and unrotated
ellipticities, and in terms of the complex ellipticity $p$,
\begin{eqnarray}
     C_1(\theta)+C_2(\theta) &=&
     \VEV{\epsilon_+^r(\vec\theta_0)\epsilon_+^r(\vec\theta_0 +
     \vec\theta)} +
     \VEV{\epsilon_\times^r(\vec\theta_0)\epsilon_\times^r(\vec\theta_0 +
     \vec\theta)} \nonumber \\
      &=& \VEV{\epsilon_+(\vec\theta_0)\epsilon_+(\vec\theta_0 +
     \vec\theta)} +
     \VEV{\epsilon_\times(\vec\theta_0)\epsilon_\times(\vec\theta_0 +
     \vec\theta)} \nonumber \\
     & = & {\rm Re}[p^*(\vec\theta_0)
     p(\vec\theta_0+\vec\theta)],
\label{pluscorrelation}
\end{eqnarray}
\begin{eqnarray}
     C_1(\theta)-C_2(\theta) &=&
     \VEV{\epsilon_+^r(\vec\theta_0)\epsilon_+^r(\vec\theta_0 +
     \vec\theta)} -
     \VEV{\epsilon_\times^r(\vec\theta_0)\epsilon_\times^r(\vec\theta_0 +
     \vec\theta)} \nonumber \\
     & =& \cos 4\phi \Biggl[
     \VEV{\epsilon_+(\vec\theta_0)\epsilon_+(\vec\theta_0 +
     \vec\theta)} \nonumber \\
     && - \VEV{\epsilon_\times(\vec\theta_0)
     \epsilon_\times(\vec\theta_0 + \vec\theta)} \Biggr]
     \nonumber \\
      && + \sin 4\phi \Biggl[ \VEV{\epsilon_\times(\vec\theta_0)
      \epsilon_+(\vec\theta_0 + \vec\theta)}  \nonumber \\
      && + \VEV{\epsilon_+(\vec\theta_0)  
     \epsilon_\times(\vec\theta_0 + \vec\theta)} \Biggr] \nonumber \\
     & =& {\rm
     Re}[p(\vec\theta_0) p(\vec\theta_0+\vec\theta)] \cos 4\phi
     \nonumber \\ 
     && + {\rm Im}[p(\vec\theta_0) p(\vec\theta_0+\vec\theta)] \sin
     4\phi,
\label{minuscorrelation}
\end{eqnarray}
where $\vec\theta=\{\theta \cos\phi,\theta \sin \phi \}$.  There
is also the third linearly independent correlation function,
\begin{eqnarray}
     C_{\rm cross}(\theta) &=&
     \VEV{\epsilon_+^r(\vec\theta_0)\epsilon_\times^r(\vec\theta_0 +
     \vec\theta)} +
     \VEV{\epsilon_\times^r(\vec\theta_0)\epsilon_+^r(\vec\theta_0 +
     \vec\theta)} \nonumber \\
     & =& - \sin 4\phi \left[
     \VEV{\epsilon_+(\vec\theta_0)\epsilon_+(\vec\theta_0 +
     \vec\theta)} - \VEV{\epsilon_\times(\vec\theta_0)
     \epsilon_\times(\vec\theta_0 + \vec\theta)} \right]
     \nonumber \\
      && + \cos 4\phi \left[ \VEV{\epsilon_\times(\vec\theta_0)
      \epsilon_+(\vec\theta_0 + \vec\theta)} + \VEV{\epsilon_+(\vec\theta_0)
     \epsilon_\times(\vec\theta_0 + \vec\theta)} \right] \nonumber \\
     & =& - {\rm
     Re}[p(\vec\theta_0) p(\vec\theta_0+\vec\theta)] \sin 4\phi
     \nonumber \\
     && + {\rm Im}[p(\vec\theta_0) p(\vec\theta_0+\vec\theta)] \cos
     4\phi,
\label{crosscorrelation}
\end{eqnarray}
and parity conservation demands $C_{\rm cross}(\theta)=0$.
Note that when written in terms of the {\it
un}rotated ellipticities or the complex ellipticity, the sum
$C_1(\theta) + C_2(\theta)$ is {\it in}dependent of $\phi$.
However, when written in terms of the unrotated ellipticities or
complex ellipticity, the difference $C_1(\theta)-C_2(\theta)$
and $C_{\rm cross}(\theta)$ {\it does} depend explicitly on
$\phi$, the relative orientation of the two points being
correlated.

The power spectra can be written in terms of the correlation
functions as
\begin{eqnarray}
     P_{gg}(\kappa) &=& {\pi \over 2} \int \theta d\theta \Biggl\{
     [C_1(\theta)+ C_2(\theta)]J_0(\kappa\theta) \nonumber \\
     && +
     [C_1(\theta)-C_2(\theta)] J_4(\kappa\theta) \Biggr\} \nonumber \\
     P_{cc}(\kappa) &=& {\pi \over 2} \int \theta d\theta \Biggl\{
     [C_1(\theta)+ C_2(\theta)]J_0(\kappa\theta) \nonumber \\
     && -
     [C_1(\theta)-C_2(\theta)] J_4(\kappa\theta) \Biggr\}.
\end{eqnarray}
Again, if the second of these is nonzero, it can only be due to
non-lensing effects, so construction of this correlation
function provides a powerful probe for the presence of
nonlensing artifacts in the data.

Of course, correlation functions $C^s(\theta)$ for the smoothed
ellipticities can be obtained by replacing ellipticities and
power spectra by the smoothed ellipticities and power spectra in
all the equations above.

\comment{
It is straightforward to construct the correlation $C_1+C_2$,
$C_1-C_2$, and $C_{\rm cross}$ with
Eqs. (\ref{pluscorrelation})--(\ref{crosscorrelation}) by
summing over all pairs of sources in bins of separation
$\theta$.  The third correlation function should be zero.  In
principle, one can then construct the power spectra from these
correlation functions.  However, this requires that the
correlation function be reliably be determined for all angular
separations.  Realistically, this is not possible:  The measured
correlation function will be unreliable at small angular
separations because of source fragmentation, and at large
angular separations because of limitations in the data.
The correlation functions will be important for
checking for systematic effects, and they may also be useful for
a maximum-likelihood analysis of the data.
}

\section{STATISTICAL ESTIMATORS}

The effects of weak lensing can be uncovered through the
measured correlation functions, power spectra, mean-square
ellipticities averaged over some given pixel size, or a full
maximum-likelihood fit to the data.  

The Fourier modes of the shear field due to weak lensing are
statistically independent.  Furthermore, if the noise map is
orientation independent, then the Fourier modes of the noise
will also be statistically independent.  Even if we use only the
simplest (although not necessarily optimal) estimator for
the power spectrum, the mean-square ellipticity, it is better
to work in Fourier space.  The predicted weak-lensing mean-square
ellipticity is due entirely to the gradient component, but
randomly oriented intrinsic source ellipticities should
contribute to the mean-square ellipticity equally through the
gradient and curl component.  The signal-to-noise ratio will
therefore be improved with a Fourier transform which allows us
to isolate the gradient and curl components.  

\subsection{Discrete Fourier Transforms and Statistical Noise}

We restrict ourselves to a survey which covers a rectangular
region of the sky.  The analysis can be extended to
irregularly-shaped regions of the sky, but only with significant
complications.  [Simple estimates of the effects of an
irregularly-shaped survey which are used for the power spectrum
of angular clustering \cite{bau94} are not easily
generalized to weak-lensing power spectra.]  We first construct
pixels of size $\theta_p \times \theta_p$ on the sky.  
This leaves us with
$N_{\rm pix}=N_x \times N_y$ pixels, where $N_x = \theta_x/\theta_p$ and
$N_y = \theta_y/\theta_p$ and $\theta_x$ and $\theta_y$ are the
dimensions of the map.  The pixels labeled by $(i,j)$ are
centered at $\vec\theta_{ij}=(i,j)\theta_p$, and
$i=0,1,...,N_x-1$ and $j=0,1,...,N_y-1$.

The ellipticities, $\epsilon_{+,ij}^{\rm obs}$ and
$\epsilon_{\times,ij}^{\rm obs}$, measured in pixel $(i,j)$ are
the sum of a weak-lensing signal and noise which arises from
intrinsic source ellipticities and measurement error,
\begin{equation}
     \epsilon_{+,ij}^{\rm obs} =      \epsilon_{+,ij}^s + \epsilon_{+,ij}^n, 
     \qquad
     \epsilon_{\times,ij}^{\rm obs} = \epsilon_{\times,ij}^s +
     \epsilon_{\times,ij}^n.
\end{equation}
We can then use an FFT to determine the $N_{\rm pix}$ Fourier
coefficients of the survey,
\begin{equation}
     \tilde \epsilon_+^{\rm obs} (\vec\kappa) = 
     \sum_{ij} \epsilon_{+,ij}^{\rm obs} e^{i
     \vec \kappa \cdot \vec \theta_{ij}},  \quad
     \tilde \epsilon_\times^{\rm obs}(\vec\kappa) = 
     \sum_{ij} \epsilon_{\times,ij}^{\rm obs} e^{i
     \vec \kappa \cdot \vec \theta_{ij}},
\end{equation}
for $\vec \kappa=(2\pi/\theta_p)(n/N_x,m/N_y)$, and
$n=0,1,...,N_x-1$ and $m=0,1,...,N_y-1$.

If the noise terms are all statistically independent with 
variances $\sigma_\epsilon^2$ [i.e., they satisfy
$\VEV{\epsilon^n_{+,ij} \epsilon^n_{+,kl}} = \sigma_\epsilon^2
\delta_{ik} \delta_{jl}$, $\VEV{\epsilon^n_{\times,ij}
\epsilon^n_{\times,kl}} =
\sigma_\epsilon^2\delta_{ik} \delta_{jl}$, and $\VEV{\epsilon^n_{+,ij}
\epsilon^n_{\times,kl}} = 0$], then estimators for
the mean-square ellipticities are given by
\begin{equation}
     \widehat{(\gamma_g^s)^2} = \left({ 1 \over N_{\rm
     pix}^2} \sum_{\vec\kappa} | \gamma_g^{\rm obs}(
     \vec\kappa)|^2 \right) - \sigma_\epsilon^2,
\end{equation}
\begin{equation}
     \widehat{(\gamma_c^s)^2} = \left( { 1 \over N_{\rm
     pix}^2} \sum_{\vec\kappa} | \gamma_c^{\rm obs}(
     \vec\kappa)|^2 \right) - \sigma_\epsilon^2,
\end{equation}
\begin{equation}
     \widehat{(\gamma_g^s)^* \gamma_c^s} = \left( { 1 \over N_{\rm
     pix}^2} \sum_{\vec\kappa} [\gamma_g^{\rm obs}(
     \vec\kappa)]^*  \gamma_c^{\rm obs}(\vec\kappa) \right) -
     P_{gc}^n(\kappa).
\end{equation}
These are estimators for variances of a distribution measured
with a finite number of pixels.  Therefore, there will be some
cosmic variance as well as some pixel-noise variance with which
these estimators will recover their expectation values.  These
variances are
\begin{equation}
     \VEV{ \left[ \widehat{(\gamma_g^s)^2} -
     \VEV{(\gamma_g^s)^2} \right]^2} = {2 \over N_{\rm
     pix}} \left[ \VEV{(\gamma_g^s)^2} +
     \sigma_\epsilon^2 \right]^2,
\label{eq:gvariance}
\end{equation}
\begin{equation}
     \VEV{ \left[ \widehat{(\gamma_c^s)^2} -
     \VEV{(\gamma_c^s)^2} \right]^2} = {2 \over N_{\rm
     pix}} \left[ \VEV{(\gamma_c^s)^2} +
     \sigma_\epsilon^2 \right]^2,
\label{eq:cvariance}
\end{equation}
\begin{eqnarray}
     \VEV{ \left[ \widehat{(\gamma_g^s)^* \gamma_g^c} -
     \VEV{(\gamma_g^s)^* \gamma_c^s} \right]^2} &=& {1
     \over N_{\rm pix}} \left[ \VEV{(\gamma_g^s)^2} +
     \sigma_\epsilon^2 \right] \nonumber \\
     && \times \left[ \VEV{(\gamma_c^s)^2} +
     \sigma_\epsilon^2 \right].
\label{eq:gcvariance}
\end{eqnarray}
These results may be obtained in analogy with the derivation for
cosmic and pixel-noise variances for a temperature-polarization
map of the cosmic microwave background \cite{kno95}.  

Inserting the null hypothesis of no signal, $\VEV{ (\gamma_g^s)^2
} =0$, into Eq. (\ref{eq:gvariance}) gives us the
statistical limit to the weak-lensing amplitude of this quantity
to which we are sensitive.  Explicitly, we can be assured a
$3\sigma$ detection of $(\gamma_g^s)^2$ only if it exceeds $3
\sigma_\epsilon^2 \sqrt{2/N_{\rm pix}}$.  

If the density of resolved sources on the sky is $\bar n$
(in units of deg$^{-2}$) and the mean intrinsic ellipticity of the
sources is $\bar\epsilon$ (the mean intrinsic
value of $|p|$), then $\sigma_\epsilon^2 =
\bar\epsilon^2 /(\bar n \theta_p^2)$.  Therefore, if the area of
the survey is $A$, then the survey will be sensitive (at
$1\sigma$) to a mean-square ellipticity,
\begin{eqnarray}
     \sigma_{\VEV{(\gamma_g^s)^2}} &=& (0.0075)^2 \,
     (A/10,000\,{\rm deg}^2)^{-1/2} \nonumber \\
     && \times (\bar\epsilon/0.4)^2 (\bar n/40\,{\rm deg}^{-2})^{-1}
     (\theta_p/{\rm deg})^{-1},
\label{eq:sigmaequation}
\end{eqnarray}
for pixels of area $\theta_p^2$.  Since the signal
is the mean-square ellipticity (rather than the rms
ellipticity), an rms ellipticity $\gap \sqrt{3}(0.0075)\simeq
0.013$ in $1^\circ$ square pixels should be detectable at
$3\sigma$ with the survey parameters assumed here.  The central
values above were chosen to be close to those expected for
resolved sources in the completed FIRST survey.

\subsection{Likelihood Analysis}

Although the mean-square ellipticity per pixel provides a simple
estimate of the sensitivity of a given survey to a signal, it
is not the optimal statistic for detecting a weak-lensing signal.
The sensitivity of a survey to a weak-lensing signal can be
improved significantly with a maximum-likelihood analysis which
compares the complete power spectrum (rather than just the
mean-square ellipticity) with the entire survey.

Suppose that in a survey with $N_{\rm pix}$ pixels we construct
a $2 N_{\rm pix}$-dimensional data vector,
$D^{\rm obs}_\alpha=\{\epsilon^{\rm obs}_{+,1}, \epsilon^{\rm
obs}_{\times,1},\epsilon^{\rm obs}_{+,2}, \epsilon^{\rm
obs}_{\times,2},...,\epsilon^{\rm
obs}_{+,N_{\rm pix}}, \epsilon^{\rm obs}_{\times,N_{\rm pix}}
\}$, from the $2N_{\rm pix}$ measured ellipticities
$\epsilon^{\rm obs}_{+,ij}$ and $\epsilon^{\rm obs}_{+,ij}$, and
each observed value is due to signal and noise, $D^{\rm
obs}_\alpha = D^s_\alpha + D^n_\alpha$.  Suppose further that
we are testing a Gaussian theory which predicts expectation
values $\VEV{ D_\alpha^s D_\beta^s}=C^s_{\alpha\beta}$
(where the theory correlation matrix is given by the unrotated
correlation functions discussed in section 2.3) with a map which
has Gaussian noise with a correlation matrix $\VEV{D^n_\alpha
D^n_\beta}=C^n_{\alpha\beta}$.  The likelihood of this theory
given the data is
\begin{equation}
     {\cal L} \propto \exp\{ D^{\rm obs}_\alpha
     [(C^s+C^n)^{-1}]_{\alpha\beta} D^{\rm obs}_\beta \}.
\label{eq:likelihood}
\end{equation}
For example, if the noise is due
entirely to intrinsic source ellipticities, then the noise in
each pixel is uncorrelated and the noise between $+$ and
$\times$ ellipticities is also uncorrelated, so the noise
correlation matrix becomes diagonal with entries equal to the
variance in each ellipticity,
$C^n_{\alpha_\beta}=\sigma_\epsilon^2 \delta_{\alpha\beta}$.
In general, however, the noise correlation matrix will be nondiagonal
(Refregier \& Brown 1997), and the theory matrix is also
nondiagonal.  Therefore, for a 10,000-deg$^2$ survey with
$20'\times 20'$ pixels, the data vector will have
180,000 entries, and evaluation of the likelihood would require
inversion of a $180,000\times180,000$ matrix!

Progress in evaluating the likelihood with good accuracy
can be made by working in the Fourier domain instead.  In this
case, we write the data as a $2 N_{\rm pix}$-dimensional
vector with the $N_{\rm pix}$ measured Fourier components
$\tilde\gamma_g^{\rm obs}(\veckappa)$ and
$\tilde\gamma_c^{\rm obs}(\kappa)$
as components.  Statistical isotropy and homogeneity guarantee
that these have expectation values
$\VEV{\tilde\gamma_g^s(\vec\kappa) \tilde
\gamma_g^s(\veckappa')} = P_{gg}^s(\kappa)
\delta_{\veckappa\veckappa'}$, $\VEV{\tilde\gamma_g^s(\vec\kappa) \tilde
\gamma_c^s(\veckappa')} = 0$, and
$\VEV{\tilde\gamma_c^s(\vec\kappa) \tilde
\gamma_c^s(\veckappa')} = 0$.  In other words, the covariance of 
the Fourier components is diagonal, and since all the $c$
components are expected to be zero, we only need to deal with a
$N_{\rm pix} \times N_{\rm pix}$ (rather than $2N_{\rm pix}
\times 2N_{\rm pix}$) covariance matrix.  Unfortunately, the
covariance of the noise in Fourier space will not always be
diagonal, unless the noise correlation functions depend only on
the separation between two points and not their relative
orientation.  This will not necessarily be the case. 
For example, the noise in interferometric images is generally
spatially correlated and has preferred directions (visible as
``stripes'' in the noise). This produces spurious ellipticity
correlations which depend on the orientation of the
source-separation vector \cite{ref97}.
Therefore, the error obtained by assuming the covariance in
Fourier space to be diagonal will underestimate the true error.
By carrying out a full likelihood analysis on a small patch of
the survey and comparing it with the results of the restricted
analysis (i.e., that which assumes the Fourier components are
all statistically independent) on that same patch, one can
determine the degree to which the restricted analysis
underestimates the error.

To proceed, we must make the approximation that the covariance
of the noise is diagonal in the Fourier domain.  Then the power
spectrum $P_{gg}^n(\kappa)$ of the noise can be obtained from
the noise correlation functions, $C_1^n(\theta)$ and
$C_2^n(\theta)$, which are obtained by averaging over all
orientations $\phi$.  With this approximation, both the noise
and theory covariance matrices are diagonal in the Fourier
domain, and evaluation of the likelihood reduces to evaluation
of the usual $\chi^2$.  Therefore, if our model consists of a
smoothed power spectrum ${\cal A} P^s_{gg}(\kappa)$ and we are
trying to fit for the amplitude ${\cal A}$, then each of the
$N_{\rm pix}$ measured Fourier components $\veckappa$ provides an
independent estimator,
\begin{equation}
     {\cal A}_\veckappa = {|\tilde\gamma_g^{\rm obs}(\veckappa)|^2 -
     P_{gg}^n(\kappa) \over P^s_{gg}(\kappa)},
\label{eq:indepestimator}
\end{equation}
for ${\cal A}$, with a variance
\begin{equation}
     \sigma_{{\cal A}_\veckappa}^2 = \VEV{ ({\cal A}_\veckappa -
     \VEV{{\cal A}_\veckappa})^2} = {2 [{\cal A}
     P^s_{gg}(\kappa) + P_{gg}^n(\kappa)]^2 \over
     [P_{gg}^s(\kappa)]^2}.
\label{indepvariance}
\end{equation}
Therefore, an estimator for ${\cal A}$ for the entire survey is
\begin{equation}
     \widehat{\cal A} = {\sum_\veckappa {\cal
     A}_\veckappa/\sigma_{{\cal A}_\veckappa}^2 \over
     1/\sigma_{\cal A}^2 },
\label{eq:totalestimator}
\end{equation}
and the variance to this estimator is $\sigma_{\cal A}$ given by
\begin{equation}
     {1 \over \sigma_{\cal A}^2} = \sum_\veckappa {1 \over
     \sigma_{{\cal A}_\veckappa}^2}.
\label{eq:totalvariance}
\end{equation}

To determine the smallest value of ${\cal A}$ which could be
distinguished from the noise, we insert the null hypothesis
${\cal A}=0$ into Eq. (\ref{eq:totalvariance}).  To illustrate,
suppose that the noise in each component of the ellipticity was
Gaussian with variance $\sigma_\epsilon^2$ and independent of
each other component, as assumed above.  In this case,
$P^n_{gg}(\kappa)=\sigma_\epsilon^2$.  Therefore,
\begin{equation}
     {1 \over \sigma_{\cal A}^2} = {1 \over
     2 \sigma_\epsilon^4} \sum_\veckappa [P_{gg}^s(\kappa)]^2
     \simeq  { N_{\rm pix}^2 \over 2
     \sigma_\epsilon^4} \int\, {d^2 \veckappa \over (2\pi)^2}
     [P_{gg}^s(\veckappa)]^2,
\end{equation}
or in other words,
\begin{equation}
     {\sigma_{\cal A} \over {\cal A}} = {\sqrt{2}
     \sigma_\epsilon^2 \over N_{\rm pix}} {\cal I}_\sigma^{-1/2}
     = {\sqrt{2} \bar\epsilon^2 \over \bar n A} {\cal
     I}_\sigma^{-1/2},
\label{eq:likelihoodresult}
\end{equation}
where
\begin{equation}
     {\cal I}_\sigma \equiv  {\cal A}^2 \int\, {d^2 \veckappa
     \over (2\pi)^2} [P_{gg}^s(\veckappa)]^2,
\end{equation}
and we have used $\sigma_\epsilon^2 = \bar\epsilon^2 /(\bar n
\theta_p^2)$ and $\theta_p^2 = A/N_{\rm pix}$.
Eq. (\ref{eq:likelihoodresult}) shows how the signal-to-noise
scales with the mean intrinsic source ellipticity
$\bar\epsilon$, usable density of sources $\bar n$ and the
survey area $A$.

\section{PREDICTIONS FOR THE FIRST SURVEY}

\subsection{Preliminaries}

We will restrict our analysis to a flat Universe
($\Omega_0+\Omega_\Lambda=1$), but will allow for a nonzero
cosmological constant $\Omega_\Lambda$.  The scale factor of the
Universe, $a(t)$, satisfies the Friedmann equations,
\begin{equation}
     {\dot a \over a} = H_0 E(z) \equiv H_0 \sqrt{\Omega_0 (1+z)^3 +
     \Omega_\Lambda},
\end{equation}
where $H_0 = 100\, h \, {\rm km}\,{\rm sec}^{-1}\,{\rm
Mpc}^{-1}$ is the Hubble constant, $\Omega_0$ is the current
nonrelativistic-matter density in units of the critical density,
$\Omega_\Lambda$ is the current contribution of the cosmological
constant to closure density, and the dot denotes derivative with
respect to time.

We choose the scale factor such that $a_0 H_0=2$.  If we are
located at the origin, $\vecw=0$, then an object at redshift $z$
is at a comoving distance,
\begin{equation}
     w(z) = {1\over 2} \int_0^z \, {dz' \over E(z')},
\end{equation}
and the comoving distance to the horizon (or the conformal time
today) is 
\begin{equation}
     \eta_0 = {1\over 2} \int_0^\infty \, {dz' \over E(z')}.
\end{equation}

\subsection{Weak-Lensing Power Spectrum}

Given a power spectrum $P(k,z)$ for the mass distribution, the power
spectrum for the gravitational potential is
\begin{equation}
     P_\phi(k,z) = k^{-4} \left[ {3 \over 2} (a_0 H_0)^2
     \Omega_0 (1+z) \right]^2 P(k,z).
\label{eq:Pphi}
\end{equation}
In linear theory, the time evolution of the power spectrum is
given by
\begin{equation}
     P(k,z) = P(k,z=0)[D(z)/D(z=0)]^2,
\label{eq:mattergrowth}
\end{equation}
where
\begin{equation}
     D(z) = {5 \Omega_0\, E(z) \over 2} \, \int_z^\infty \, {1+z' \over
     [E(z')]^3} \, dz',
\label{eq:Deqn}
\end{equation}
is the linear-theory growth factor (see, e.g., Peebles 1993).
     
The weak-lensing power spectrum is \cite{kai92}
\begin{equation}
     P_{gg}(\kappa) = 4 \int\, dz\, {dw \over dz} \left[ {g(z) \over w(z)
     } \right]^2 \left[ {\kappa \over w(z) } \right]^4 \, P_\phi(\kappa/w,z),
\label{eq:Peq}
\end{equation}
where $g(z)$ is given in terms of the survey redshift distribution,
$dN/dz$, by
\begin{equation}
     { g(z) \over w(z) } = \int_z^\infty \, dz'\, {1\over N}{ dN \over dz'} \,
     \left[ 1 - {w(z) \over w(z')} \right].
\end{equation}
Inserting Eqs. (\ref{eq:Pphi})--(\ref{eq:Deqn}) into
Eq. (\ref{eq:Peq}), we find
\begin{equation}
     P_{gg}(\kappa) = {1 \over \kappa} \,
     \int_{\kappa/\eta_0}^\infty \, I_\kappa(k) \, P(k,z=0)\, dk,
\end{equation}
where
\begin{eqnarray}
     I_\kappa(k) &=& 144 \Omega_0^2 [g(w=\kappa/k)]^2
     [1+z(w=\kappa/k)]^2 \nonumber \\
     && \times [D(w=\kappa/k)/D_0]^2.
\end{eqnarray}
The mean-square variance in a cell with window function $\widetilde
W(\kappa)$ is
\begin{equation}
     \VEV{ (\gamma_g^s)^2 } = \int\, dk \, P(k,0)\, G(k),
\label{eq:predict}
\end{equation}
where
\begin{equation}
     G(k) = \int_0^{k\eta_0} \, {d\kappa \over 2 \pi} |\widetilde
     W(\kappa)|^2 \, I_\kappa(k).
\end{equation}

\subsection{Model for the Spatial Density Power Spectrum}

For the power spectrum, we use
\begin{eqnarray}
     P(k) &=& {2 \pi^2 \over 8} \delta_H^2 (k/2)^n T^2(k_p\,{\rm
     Mpc}/h \Gamma),
\end{eqnarray}
where $T(q)$ is the usual CDM transfer function, $k_p= k
H_0/2$ is the physical wave number, and $\Gamma \simeq \Omega_0
h$ is given more accurately in terms of $\Omega_0 h$ and the
baryon fraction $\Omega_b$ by Eqs. (D-28) and (E-12) in Hu \&
Sugiyama (1996).  For the transfer function, we use \cite{bar86},
\begin{equation}
     T(q)= {\ln(1+2.34q)/(2.34 q)\over [1+3.89 q + (16.1 q)^2 +
     (5.46 q)^3 + (6.71 q)^4]^{1/4}}.
\label{transferfunction}
\end{equation}

If the power spectrum is normalized to {\sl COBE}, the
amplitude $\delta_H$ is \cite{bun97}
\begin{eqnarray}
     \delta_H(n,\Omega_0) &=& 1.94\times 10^{-5}\,
     \Omega_0^{-0.785-0.05\,\ln\Omega_0} \nonumber \\
     && \times \exp[a (n-1) +
     b(n-1)^2].
\end{eqnarray}
If primordial density perturbations are due to inflation, then
there will also be a stochastic gravity-wave background which
contributes to the {\sl COBE} anisotropy with an amplitude
dependent on the spectral index $n$.  In this case, $a=1$ and
$b=1.97$.  If we make no such assumption and suppose that the
stochastic gravity-wave background is negligible, then $a=-0.95$
and $b=-0.169$.  If we are uncertain of the gravity-wave
contribution to {\sl COBE}, then the {\sl COBE} normalization
above (with no gravity-wave background) will provide an upper
limit to the true amplitude of the power spectrum.

Alternatively, the power
spectrum may be normalized at small distance scales through the
cluster abundance which fixes $\sigma_8$, the variance in the
mass enclosed in spheres of radius $8\,h^{-1}$~Mpc, to
$\sigma_8\simeq (0.6 \pm 0.1)\Omega_0^{-0.53}$ \cite{via96}.  In
terms of the power spectrum,
\begin{equation}
     \sigma_8^2 = {1 \over 2 \pi^2} \int\,k^2 \, dk \, P(k)
     \left[ { 3 j_1(k_p R) \over k_p R} \right]^2,
\end{equation}
where $R=8\,h^{-1}$ Mpc, and $j_1(x)$ is a spherical Bessel
function.  Since we are using $a_0\neq1$, $k_p$ (rather than
$k$) enters into the argument of the spherical Bessel function.

\subsection{Results for the FIRST Survey}

\begin{figure}
\epsfxsize=85truemm 
\epsffile{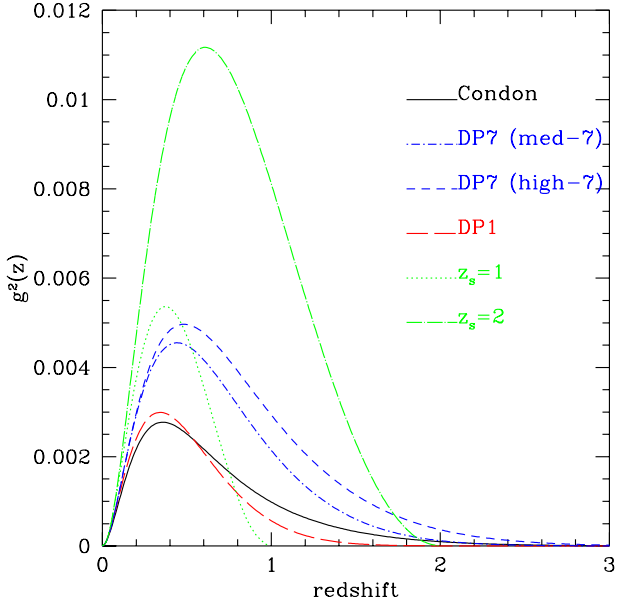}
\caption[fig:zplot]{Redshift weight functions $g^2(z)$ for the four {\sl
     FIRST} redshift distributions considered in Cress \&
     Kamionkowski (1997).  Also shown are $g^2(z)$ that would be
     obtained if all the sources were at redshift $z_s=1$ or
     $z_s=2$.}
\label{fig:zplot}
\end{figure}

Fig. \ref{fig:zplot} shows the weight function $g^2(z)$ that
enters into the calculation of the weak-lensing signal for the
four {\sl FIRST} redshift distributions considered by Cress \&
Kamionkowski (1997) and shown in Fig. 1 therein.  Our best
estimate for the {\sl FIRST} redshift distribution is ``DP7
(med-$z$)'' derived from a radio-source luminosity function due to
Dunlop and Peacock (1990), but we also include two other
plausible estimates from these authors, ``DP7 (high-$z$)'' and
``DP1,'' as well as a redshift distribution derived
{}from a luminosity function due to Condon (1984).  For
comparison, we also show the weight functions obtained from
assuming all sources to be at a redshift of $z_s=1$ or $z_s=2$.
Below, we calculate the predicted signals with all four {\sl
FIRST} redshift distributions to assess the uncertainty in the
predictions from imprecise knowledge of the redshift
distribution.

\begin{figure}
\epsfxsize=85truemm 
\epsffile{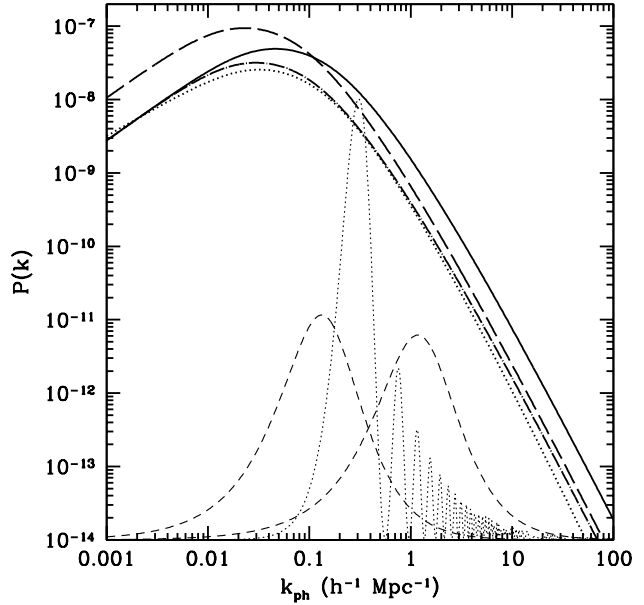}
\caption[fig:Pplot]{Three-dimensional {\sl COBE}-normalized power spectra
     $P(k)$ for the four models listed in Table 1.  Models 1--4
     are represented by solid, dotted, dash, and dot-dash
     curves, respectively.  Also shown (the light dashed
     curves) are the window functions $G(k)$ needed for
     calculation of the mean-square ellipticity at
     $\theta_p=1^\circ$ and $\theta_p=6'$, and the window
     function (light dotted curve) needed for the calculation of
     $\sigma_8$.}
\label{fig:Pplot}
\end{figure}

\begin{figure}
\epsfxsize=85truemm 
\epsffile{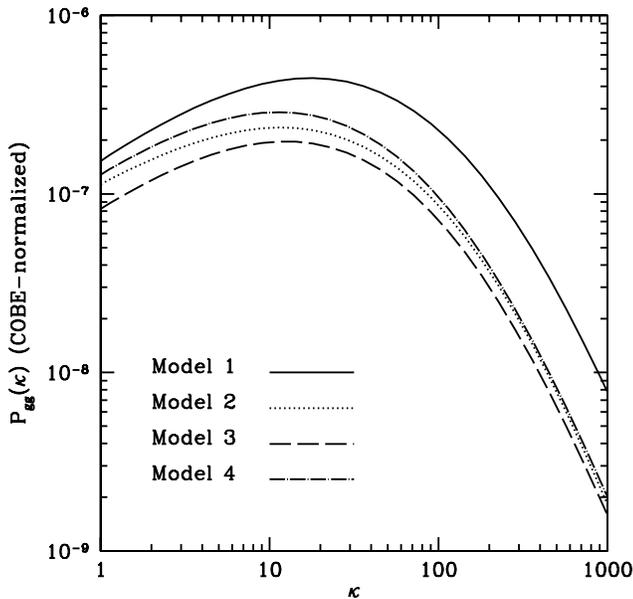}
\caption[fig:Pggplot]{Weak-lensing power spectra $P_{gg}(\kappa)$ for the
     four ({\sl COBE}-normalized) models listed in Table 1.}
\label{fig:Pggplot}
\end{figure}

\begin{figure}
\epsfxsize=85truemm 
\epsffile{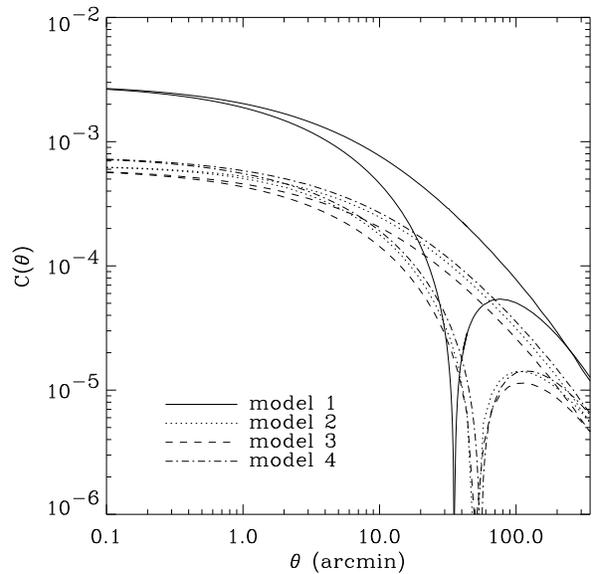}
\caption[fig:correlations]{Weak-lensing correlation functions $C_1(\theta)$ and
     $C_2(\theta)$ for the four ({\sl COBE}-normalized) models
     listed in Table 1.}
\label{fig:correlations}
\end{figure}

Fig. \ref{fig:Pplot} shows the spatial power spectra of the mass
distribution for the four models listed in Table
\ref{tab:predictions}.  The light dashed curves are the
(unnormalized) window functions $G(k)$ needed to calculate the mean-square
ellipticity for $1^\circ\times1^\circ$ square pixels and
$6'\times6'$ square pixels (obtained with our best
estimate, DP7 (med-$z$), for the {\sl FIRST} redshift distribution).  The
light dotted curve is the window function needed to calculate
$\sigma_8$. These window functions illustrate that weak lensing probes
power over a wide range of distance scales, and that the
ellipticity in $1^\circ\times1^\circ$ pixels probes the power
spectrum on larger scales than $\sigma_8$.  Here we used the DP7
(med-$z$) redshift distribution.  The weak-lensing window
functions would be shifted very slightly to larger scales if we
had used the high-$z$ DP7 redshift distribution and to slightly
smaller scales if we had used the DP1 or Condon redshift
distributions.  Figs. \ref{fig:Pggplot} and
\ref{fig:correlations} show the (unsmoothed) weak-lensing power
spectra and correlation functions for the four ({\sl
COBE}-normalized) models listed in Table \ref{tab:predictions}
(again, using the med-$z$ DP7 redshift distribution).

\begin{table*}
\caption{Predicted RMS ellipticities,
  $\VEV{(\gamma_g^s)^2}^{1/2}$ of {\sl FIRST}
  radio sources from weak lensing for $\theta_p=6'$, 
  $20'$, and $1^\circ$ for both {\sl COBE}-normalized and
  cluster-abundance--normalized power spectra.  We also list the
  values of $\sigma_8$ for {\sl COBE}-normalized models.}
\begin{center}
\begin{tabular}{|c|c|c|c|c|c|c|c|c|c|c|c|}
  \multicolumn{6}{c}{} &
  \multicolumn{6}{c}{($100\,\VEV{(\gamma_g^s)^2}^{1/2}$)}
  \\ 
  \multicolumn{6}{c}{} & \multicolumn{3}{c}{{\sl
  COBE}-normalized} & 
      \multicolumn{3}{c}{cluster normalized} \\ \hline
 $\Omega$ & $h$ & $n$ & $\Omega_b h^2$ & $\sigma_8$ & $\sigma_8
 \Omega_0^{0.53}$ & $6'$ &
 $20'$ & $1^\circ $ &  $6'$ &
 $20'$ & $1^\circ $  \\ \hline \hline
 1 & 0.5 & 1 & 0.0125 & 1.21 & 1.21 & 5.2 & 3.7 & 2.2 & 2.6 & 
       1.8 & 1.1\\ \hline
 1 & 0.5 & 0.8 & 0.025 & 0.71 & 0.71 & 2.8   & 2.1 &  1.4 & 2.4 
      & 1.8 & 1.2\\ \hline
 0.4 & 0.65 & 1 & 0.015 & 1.07 & 0.65 & 2.6 & 1.9 & 1.3 & 2.4 &
      1.8 & 1.2 \\ \hline
 1 & 0.35 & 1 & 0.015 & 0.74 & 0.74 & 2.9 & 2.2 & 1.5 & 2.4 &
      1.8 & 1.2\\ \hline 
\end{tabular}
\end{center}
\label{tab:predictions}
\end{table*}

Table \ref{tab:predictions} lists the predicted
rms gradient component of the
ellipticity for several flat {\sl COBE}-normalized and
cluster-abundance--normalized cold-dark-matter models with
and without a cosmological constant for $6' \times 6'$ pixels, $30'
\times 30'$ pixels, and $1^\circ \times 1^\circ$ pixels calculated with
Eq. (\ref{eq:predict}).  We used the med-$z$ DP7
redshift distribution for these calculations.  For the window
function, we use
\begin{equation}
     \widetilde W(\kappa) = 2 J_1(x)/x \qquad {\rm with} \qquad
     x\equiv \kappa\theta_p/\sqrt{\pi},
\label{eq:windowfunction}
\end{equation}
and $\theta_p$ is the pixel size.\footnote{Actually, this is the window
function for circular pixels of the same area.  The results
should be similar if we use the window function for square
pixels.}  In all cases, the rms ellipticity scales with the
pixel size roughly as $\VEV{(\gamma_g^s)^2}^{1/2}
\propto\theta_p^{-\beta}$ with $\beta=0.3-0.4$.  We also list $\sigma_8
\Omega_0^{0.53}$ where $\sigma_8$ is that obtained when the
power spectrum is normalized to {\sl COBE}.  Note that if the
power spectrum is normalized to $\sigma_8 \Omega_0^{0.53}=0.6$,
as indicated by the cluster abundance, then the rms
ellipticities are $0.018$ and $0.012$ for $\theta_p=20'$
and $1^\circ$, independent of the model (except for the
$\Omega_0=1$, $h=0.5$ model which differs negligibly).  Weak
lensing is due to perturbations in the gravitational potential
(rather than the mass distribution), and the amplitude of
gravitational-potential perturbations is fixed by the cluster
abundance.  This is why the weak-lensing power spectrum and
correlations functions shown in Figs. \ref{fig:Pggplot} and
\ref{fig:correlations} for the {\sl COBE}-normalized Model 1 are
so much higher than the others: this model predicts a value of
$\sigma_8 \Omega_0^{0.53}$ significantly larger than the others.

We have checked that our calculations agree reasonably well with
those of Jain \& Seljak (1997) for the models and redshift
distributions they consider.  Their work illustrates that
nonlinear effects (which
we have not taken into account) are important only for
$\theta_p\lap 10'$.  Corrections to the predicted signal
due to nonlinear evolution of the power spectrum should increase
the weak-lensing signal for $\theta_p \gap 10'$, but only by
a relatively small amount.

\begin{table*}
\caption{Predicted RMS ellipticities (in percent) of radio
  sources from weak lensing for the {\sl COBE}-normalized Model
  1 for $06'\times6'$ and $1^\circ\times1^\circ$ pixels.}
\begin{center}
\begin{tabular}{|c|c|c|}
\hline
 Redshift Distribution &
 $100\,\VEV{(\gamma_g^s)^2}^{1/2}(\theta_p = 6')$ & 
 $100\,\VEV{(\gamma_g^s)^2}^{1/2}(\theta_p=1^\circ)$\\ \hline \hline
 DP7 (med-$z$) & 5.2 & 2.2 \\ \hline
 DP7 (high-$z$) & 5.6 & 2.3 \\ \hline
 DP1 & 4.0 & 1.8 \\ \hline
 Condon & 4.2 & 1.9 \\ \hline
\end{tabular}
\end{center}
\label{tab:comparison}
\end{table*}

Table \ref{tab:comparison} lists the predictions for the {\sl
COBE}-normalized $\Omega_0=1$, $h=0.5$, and $n=1$ model for the
four {\sl FIRST} redshift distributions.  If the high-$z$ DP7
redshift distribution is adopted, rather than the med-$z$ DP7
distribution, then the predicted weak-lensing signal is
increased by about 6\%.  But if the true redshift distribution
were more accurately represented by the DP1 or Condon
distribution, the signal would be smaller by about 20--25\%.
Note that the fractional uncertainty in the weak-lensing signal
is smaller for larger pixel sizes than it is for smaller pixel
sizes.

\subsection{Detectability of a Signal with FIRST}

Table \ref{tab:predictions} shows, for example, that for a {\sl
COBE}-normalized CDM power spectrum with $\Omega_0=1$ and
$h=0.5$, the  predicted mean-square ellipticity in $1^\circ$
square pixels is $(0.022)^2$ and $(0.037)^2$ in $20'$
square pixels.  The rms noise in $1^\circ$ for the survey
parameters used above ($A=10,000$ deg$^2$, $\bar n=40$
deg$^{-2}$, and $\bar\epsilon=0.4$) is $(0.0075)^2$ and it 
is $(0.015)^2$ for $20'$ square pixels, which gives
signal-to-noise ratios of 9 and 6 for $1^\circ$ and
$20'$ pixels, respectively.  If the power spectrum is
normalized to the cluster abundance, then the signal is just
near the detection threshold.  Since the signal increases with
smaller pixel size as $\theta_p^{-2\beta}$ (with
$\beta=0.2-0.3$) and the noise increases as $\theta_p^{-1}$, the
sensitivity decreases slightly with if a smaller smoothing scale
is chosen.

However, the sensitivity of the signal can be improved
significantly if the full information encoded in the power
spectrum is exploited with a maximum-likelihood analysis.
For example, for the {\sl COBE}-normalized $\Omega_0=1$ and
$h=0.5$ model, ${\cal I}_\sigma=1.6\times10^{-10}$ for the
window function corresponding to $20'$ pixels.  From
Eq. (\ref{eq:likelihoodresult}), the
signal-to-noise with this maximum-likelihood technique would be
${\cal A}/\sigma_{\cal A}=22$, which is much larger than that
obtained by just comparing the predicted and measured
mean-square ellipticity using either $20'$ or $1^\circ$
pixels.  Therefore, a proper maximum-likelihood analysis can
improve the sensitivity by a factor of 2--3.  Given that the
signal for cluster-abundance--normalized power spectra is only
on the verge of detectability when only the
mean-square ellipticity is measured (i.e., the predicted mean-square
ellipticity is only slightly larger than $3\sigma$), we conclude
that, with this more sophisticated maximum-likelihood analysis, a
high-significance detection ($\gap 6\sigma$) should be
possible with these survey parameters.

\section{DISCUSSION AND CONCLUSIONS}

In this paper we have calculated the predicted ellipticity
correlations of {\sl FIRST} radio sources expected from weak
gravitational lensing due to mass inhomogeneities along the line
of sight for several plausible power spectra for the large-scale
mass distribution in the Universe.  We discussed the tensor
Fourier analysis and statistical techniques needed to isolate
the signal in the data. The shear field reconstructed from
measured ellipticities can be decomposed into a ``gradient'' and
``curl'' component.  Weak lensing predicts the presence of only
a gradient component.  Measurement of the curl component can
be used to look for non-lensing artifacts in the data.

We also estimated the amplitude of a signal which could be
detectable with a survey as a function of the survey's source
density, mean intrinsic source ellipticity, and area of the
survey.  We found that a detection of the
signal from cluster-abundance--normalized power spectra could be
expected with good statistical significance ($\gap 6\sigma$)
with survey parameters which approximate those of {\sl FIRST}.
{\sl COBE}-normalized models produce an even larger signal.

In addition to the statistical errors which we have taken into
account, there will be systematic effects in the data which will
mimic the effects of weak lensing.  However, the most egregious
of these effects can be modeled and corrected for
\cite{ref97,refetal97} and it should be possible to approach
the statistical limits discussed here.  Even with a
slight degradation of the signal-to-noise expected from
systematic effects, the effects of weak lensing should be
visible for cluster-abundance--normalized power spectra with a
maximum-likelihood analysis.  A null result would place strict
upper limits on the amplitude of mass (rather than
luminous-matter) inhomogeneities in the Universe.  Seljak (1997)
has recently discussed application of more sophisticated
statistical techniques developed primarily for clustering and
the cosmic microwave background to weak lensing from large-scale
structure.  These will be needed for precise determination
of the power spectrum for future weak-lensing surveys with
better sensitivity.

Although there are several other searches for weak-lensing
correlations with optical surveys (e.g., Mould et al. 1994), as well
as some recent claimed detections \cite{vil95,sch97},
these optical surveys probe the ellipticity correlation function
on much smaller angular scales than {\sl FIRST}, which will
probe the correlation function on scales $\gap 1^\circ$.
Therefore, by combining the results of these surveys, the
weak-lensing power spectrum can be reconstructed over a wide
angular range.  Although the signal may be more easily detected
with optical surveys, these will probe scales where corrections
due to nonlinear evolution of the power spectrum may be
significant.  On the other hand, {\sl FIRST} will probe the
power spectrum in a regime where nonlinear effects are small, so
the comparison with theory will be less hampered by
theoretical uncertainties from nonlinear effects.  

The Sloan Digital Sky Survey (SDSS) will provide yet another
data base with which to look for the effects of weak lensing on
large angular scales \cite{ste96}.  However, SDSS sources
will typically be at smaller redshifts.  Therefore, by comparing
results from the SDSS and {\sl FIRST}, the redshift distribution
of the weak-lensing distortions can be disentangled.  Since {\sl
FIRST} and the SDSS will cover the same region of the sky, one
can also cross-correlate the shear field indicated by {\sl
FIRST} with the foreground density field mapped by the SDSS.
This will provide more stringent probes of the power spectrum
and should also allow a direct measurement of the bias of SDSS
sources.

\section*{Acknowledgments}

We thank S. Brown, D. Helfand, N. Kaiser, and G. Lewis for
useful discussions.  This work was supported at Columbia by
D.O.E. contract DEFG02-92-ER 40699, NASA NAG5-3091, NSF
AST94-19906, and the Alfred P. Sloan Foundation. A.B. gratefully
acknowledges financial support from New York University and
University of Victoria, and through an operating grant from
NSERC.  A.R. was supported at Princeton by the MAP/MIDEX
project.

{}


\begin{thebibliography}{}

\bibitem[Baugh \& Efsthathiou 1994]{bau94} Baugh C. M.,
     Efstathiou G., 1993, MNRAS, 265, 145
\bibitem[Bardeen et al. 1986]{bar86} Bardeen J. M., Bond
     J. R., Kaiser N., Szalay A. S., 1986, ApJ, 304, 15
\bibitem[Bartelmann \& Schneider 1992]{bar92} Bartelmann M.,
     Schneider P., 1992, A\&A, 259, 413 
\bibitem[Becker et al. 1995]{bec95} Becker R. H., White R. L.,
     Helfand D. J., 1995, ApJ, 450, 559
\bibitem[Blandford et al. 1991]{bla91}  Blandford R. D., Saust
     A. B., Brainerd T. G., Villumsen J. V., 1991, MNRAS,
     251, 600
\bibitem[Bunn \& White 1997]{bun97} Bunn E. F., White
     M., 1997, ApJ, 480, 6
\bibitem[Condon 1984]{con84} Condon J. J., 1984, ApJ, 287, 461
\bibitem[Cress \& Kamionkowski 1997]{cre97} Cress C. M.,
     Kamionkowski M., 1997, MNRAS, submitted
\bibitem[Dunlop \& Peacock 1990]{dun90} Dunlop J. S.,
     Peacock J. A., 1990, MNRAS, 247, 19 
\bibitem[Gunn 1967]{gun67} Gunn J. E., 1967, ApJ, 150, 737
\bibitem[Hu \& Sugiyama 1996]{hu96} Hu W., Sugiyama, N., 1996, ApJ,
     471, 542
\bibitem[Jain \& Seljak 1997]{jai97} Jain B., Seljak
     U., 1997, astro-ph/9611077, ApJ (in press)
\bibitem[Kaiser 1992]{kai92} Kaiser N., 1992, ApJ, 388, 272
\bibitem[Kaiser 1996]{kai96} Kaiser N., 1996, astro-ph/9610120
\bibitem[Kamionkowski et al. 1997]{kam97} Kamionkowski M.,
     Kosowsky A., Stebbins A., 1997, Phys. Rev. D, 55, 7368
\bibitem[Knox 1995]{kno95} Knox L., 1995, Phys. Rev. D, 52, 4307
\bibitem[Miralda-Escud\'e 1991]{mir91} Miralda-Escud\'e J., 1991,
     ApJ, 380, 1
\bibitem[Mould et al. 1994]{mou94} Mould J. et al., 1994, MNRAS,
     271, 31
\bibitem[Peebles 1993]{pee93} Peebles P. J. E., 1993, Principles
     of Physical Cosmology (Princeton University Press, Princeton)
\bibitem[Refregier \& Brown 1997]{ref97} Refregier A., Brown
     S. T., 1997, in preparation
\bibitem[Refregier et al. 1997]{refetal97} Refregier A.,
     Babul A., Becker R. H., Brown S. T., Cress C. M.,
     Helfand D. J., Kamionkowski M., White R. L., 1997, in
     preparation
\bibitem[Seljak 1997]{sel97} Seljak U., 1997, astro-ph/9711124
\bibitem[Schneider et al. 1997]{sch97} Schneider P., van
     Waerbeke L., Mellier Y., Jain B., Seitz S., Fort
     B., 1997, astro-ph/9705122
\bibitem[Stebbins et al. 1996]{ste96} Stebbins A., McKay T.,
     Frieman J., 1996, in Astrophysical Applications of
     Gravitational Lensing, ed. C. S. Kochanek, J. N. Hewitt
     (Netherlands: IAU), 75
\bibitem[Stebbins 1997]{ste97} Stebbins A., 1997, ApJ, in press
\bibitem[Viana \& Liddle 1996]{via96} Viana P. T. P., Liddle
     A., 1996 MNRAS, 281, 323
\bibitem[Villumsen 1995]{vil95} Villumsen J. V., 1995,
     astro-ph/9507007
\bibitem[Villumsen 1996]{vil96} Villumsen J. V., 1996, MNRAS,
     281, 369
\bibitem[White et al. 1997]{whi97} White R. L., Becker R. H.,
     Helfand D. J., Gregg M.D., 1997, ApJ, 475, 479

\end{thebibliography}
\end{document}